\documentclass{article}

\usepackage{microtype}
\usepackage{graphicx}
\usepackage{subfigure}
\usepackage{booktabs} 

\usepackage{hyperref}
\usepackage{rotating}

\usepackage[accepted]{icml2023}

\usepackage{amsmath}
\usepackage{amssymb}
\usepackage{mathtools}
\usepackage{amsthm}

\usepackage[capitalize,noabbrev]{cleveref}

\usepackage{bm}

\icmltitlerunning{
\sys: Generating musical accompaniments from singing
}

\newcommand{\sys}{\texttt{SingSong}}
\newcommand{\audiolm}{AudioLM}
\newcommand{\soundstream}{SoundStream}
\newcommand{\wtvbert}{w2v-BERT}
\newcommand{\mdxnet}{MDXNet}
\newcommand{\noisewav}{\bm{z}}
\newcommand{\voxwav}{\bm{x}}
\newcommand{\inswav}{\bm{y}}
\newcommand{\somewav}{\bm{w}}
\newcommand{\voxcodes}{\bm{\hat{x}}}
\newcommand{\inscodes}{\bm{\hat{y}}}
\newcommand{\somecodes}{\bm{\hat{w}}}
\newcommand{\enc}{\texttt{Enc}}
\newcommand{\dec}{\texttt{Dec}}
\newcommand{\semenc}{\texttt{Sem}}
\newcommand{\acoenc}{\texttt{Enc}}
\newcommand{\acodec}{\texttt{Dec}}
\newcommand{\feats}{\texttt{Feats}}
\newcommand{\acocoarse}[1]{\texttt{Coarse}(#1)}
\newcommand{\acofine}[1]{\texttt{Fine}(#1)}
\newcommand{\sample}[1]{#1\prime}
\newcommand{\fad}{FAD}
\newcommand{\fadi}{$\text{FAD}_{\text{i}}$}
\newcommand{\fads}{$\text{FAD}_{\text{s}}$}
\newcommand{\nll}{NLL}
\newcommand{\nlli}{$\text{NLL}_{\text{i}}$}
\newcommand{\nlls}{$\text{NLL}_{\text{s}}$}
\newcommand{\tabonerow}[7]{#1 & #3 & #4 & #5 \\}
\newcommand{\tabapprow}[7]{#1 & #2 & #3 & #4 & #5 & #6 & #7 \\}

\newcommand{\units}[2]{$#1$#2}
\newcommand{\CI}{\mathrel{\perp\mspace{-10mu}\perp}}
\newcommand{\expname}[1]{\texttt{#1}}

\begin{document}

\twocolumn[
\icmltitle{
\sys: Generating musical accompaniments from singing
}



\icmlsetsymbol{equal}{*}

\begin{icmlauthorlist}




\icmlauthor{Chris Donahue}{equal}
\icmlauthor{Antoine Caillon}{equal,ircam}
\icmlauthor{Adam Roberts}{equal}
\\
\icmlauthor{Ethan Manilow}{}
\icmlauthor{Philippe Esling}{ircam}
\icmlauthor{Andrea Agostinelli}{}
\icmlauthor{Mauro Verzetti}{}
\icmlauthor{Ian Simon}{}
\icmlauthor{Olivier Pietquin}{}
\icmlauthor{Neil Zeghidour}{}
\icmlauthor{Jesse Engel}{}

\vspace{1em}

Google Research

\end{icmlauthorlist}

\icmlaffiliation{ircam}{IRCAM - Sorbonne Universit\'e (work done while visiting Google Research)}

\icmlcorrespondingauthor{Jesse Engel}{jesseengel@google.com}

\icmlkeywords{Music, audio, generative models, applications}

\vskip 0.3in
]



\printAffiliationsAndNotice{\icmlEqualContribution} 

\begin{abstract}
We present \sys, a system that generates instrumental music to accompany input vocals, 
potentially offering musicians and non-musicians alike an intuitive new way to create music featuring their own voice. 
To accomplish this, we build on recent developments in musical source separation and audio generation. 
Specifically, we apply a state-of-the-art source separation algorithm to a large corpus of music audio to produce aligned pairs of vocals and instrumental sources. 
Then, we adapt 
\audiolm~\cite{borsos2022audiolm}---a state-of-the-art approach for unconditional audio generation---to
be suitable for conditional ``audio-to-audio'' generation tasks, and train it on the source-separated (vocal, instrumental) pairs. 
In a pairwise comparison with the same vocal inputs, listeners expressed a significant preference for instrumentals generated by \sys{} compared to those from a strong retrieval baseline.
\end{abstract}

\section{Introduction}

Singing is among the most intuitive pathways for engaging with music and is accessible to musicians and non-musicians alike. 
Though singing along to \emph{existing} music is a common activity, 
singing might also constitute an intuitive control mechanism for music generation systems, 
potentially allowing anyone who can sing to \emph{create} music in a playful and participatory fashion. 

\begin{figure*}[t]
\begin{center}
\centerline{\includegraphics[width=0.99\linewidth]{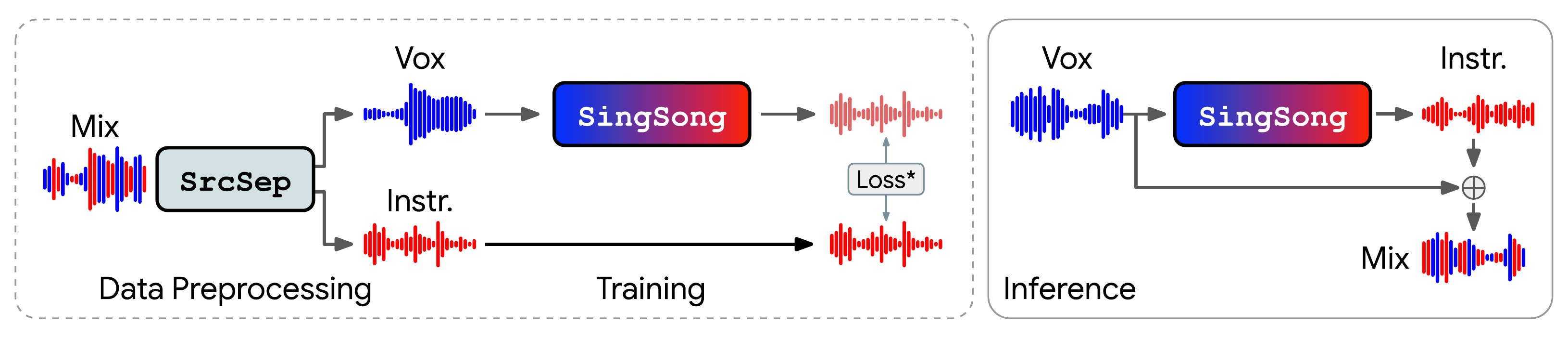}}
\caption{
\sys{} generates instrumental music to accompany input vocals, thereby allowing users to create music featuring their own voice. 
(\textbf{Left})~We manufacture large volumes of synthetic data for this task by applying an off-the-shelf source separation algorithm to a large corpus of music audio, which we use to train a generative model over instrumentals given vocals. 
(\textbf{Right})~At inference time, \sys{} takes vocals from users and outputs an instrumental to accompany to those vocals, which can be naively mixed with the input to create coherent music. 
*Note that, during training, we compute loss on discrete audio features rather than waveforms (\Cref{sec:audiotoaudiolm}). 
}
\label{fig:one}
\end{center}
\vspace{-2em}
\end{figure*}

In this work, we present \sys{} (\Cref{fig:one}), a system capable of generating instrumental music audio to accompany input vocal audio in \emph{lockstep}, i.e.,~the output instrumental can be naively mixed with the input vocals to create coherent music featuring the input. 
\sys{} leverages improvements in two key areas of music technology: source separation, and generative modeling of audio. 
We use an off-the-shelf source separation algorithm from \citet{kim2021mdx} to separate a large, diverse corpus of music ($1$M tracks) into aligned pairs of vocals and instrumental sources, constituting parallel data for our task. 
We then adapt 
AudioLM~\citep{borsos2022audiolm}---an 
unconditional generative model of 
audio involving a hierarchy of intermediate representations---to
be suitable for conditional ``audio-to-audio`` generative modeling of instrumentals given vocals, and train it on the source-separated data in a supervised fashion. 

A key challenge in this work was building a system capable of generalizing from the source-separated vocal inputs observed during training to real-world, \emph{isolated} vocals one would anticipate from users engaging with the system. 
Preliminary experiments 
resulted in models that 
had a strong bias towards reconstructing 
the instrumental from its barely-audible \emph{artifacts} present in the source-separated vocals---these models produced nonsensical outputs when fed isolated vocals. 
To improve generalization, we propose two featurization strategies for the input vocals: 
(1)~adding noise to vocal inputs to conceal artifacts, 
and 
(2)~using only the coarsest intermediate representations from \audiolm{} as conditioning input. 
On a perceptually-relevant metric, these featurizations together improve performance on isolated vocals by $55$\% relative to the default \audiolm{} featurization. 

In a pairwise study where listeners were presented two vocal-instrumental mixes with the same vocals, 
listeners expressed a significant preference for instrumentals 
from
\sys{} compared to to those from a strong retrieval baseline.\footnote{Examples: \url{https://g.co/magenta/singsong}} 
This baseline uses musical features of the vocals (beats and key) as a query to retrieve human-composed instrumentals with similar features.
Compared to instrumentals from this retrieval baseline, 
listeners preferred instrumentals from \sys{} $66$\% of the time. 
Moreover, even when comparing to ground truth instrumentals, listeners preferred instrumentals from \sys{} $34$\% of the time. 

\newpage

In summary, our primary contributions are:

\begin{itemize}
    \item We are the first to use generative modeling to create coherent instrumental accompaniments for vocal inputs.
    \item We are the first to propose using source separation to create training data for audio-to-audio accompaniment.
    \item We adapt a state-of-the-art unconditional audio generative model for conditional, audio-to-audio modeling.
\end{itemize}

\section{Related work}

\paragraph{Accompaniment generation.}

The most similar work to our own is a commercial product called Microsoft Songsmith, based on~\citet{simon2008mysong}. 
Songsmith extracts pitch information from input vocals 
and 
predicts a sequence of symbolic chord labels suitable for that melody. 
The predicted chords are then used as a query to retrieve a suitable human-composed symbolic instrumental accompaniment, which is mixed with the input vocals.
We directly compare to a similar retrieval baseline that we construct.
\citet{lattner2019high} generate symbolic kick drum accompaniments, and \citet{grachten2020bassnet} generate bass accompaniments. 
\citet{wu2022jukedrummer} generate audio of drum tracks given music audio without drums. 
Here we consider the broader task of generating the entire accompaniment track (which may contain drums, bass, and other harmonic elements) based solely on a vocal performance. 
There is also a body of work around symbolic harmonization, the task of predicting chord labels appropriate for a symbolic melody input~\citep{Paiement2006ProbabilisticMH,Yeh2020}. 
To be suitable for vocal accompaniment, such methods would require extra steps of vocal transcription and chord arrangement.

\paragraph{Music audio generation.}

Music audio generation is a challenging task, as it involves the modeling of many different audio scales, 
from 
local
(e.g.,~timbre, transients) to 
global 
information (e.g.,~musical structure, genre). 
Some work on unconditional music generation operates directly on time-domain music waveforms~\citep{oord2016wavenet,Mehri2017,donahue2018adversarial,goel2022sashimi}. 
These approaches are limited to modeling narrow distributions of music audio (commonly, piano music). 
\citet{van2017neural} introduced a method for generative modeling of invertible discrete representations of music audio. 
\citet{dieleman2018challenge} models hierarchical discrete features of piano music audio---a similar approach was used by~\citet{dhariwal2020jukebox} to create Jukebox, the first generative model of broad music audio. 
\citet{hawthorne2022general,borsos2022audiolm} also model discrete music featurizations of piano music. 
A large body of work centers around controllable music audio generation conditioned on 
symbolic representations~\citep{hawthorne2018enabling,engel2019gansynth,Hawthorne2022}, 
or performance characteristics~\citep{Caillon2021,engel2020ddsp}.
In our work, we seek to model broad distributions of instrumental music audio conditioned on vocal audio.

Contemporaneous to this work,~\citet{agostinelli2023musiclm} propose MusicLM which adapts \audiolm{} to generate broad music audio conditioned on input text descriptions. 
MusicLM is also capable of generating music with melodies corresponding to melodic \emph{hints} from users, expressed via singing or humming. 
Here our goal is to generate instrumental music in lockstep with input vocals such that they can be mixed together coherently---not only are the vocals used to steer music generation as in MusicLM, they are also directly \emph{featured} in the output music.

\section{Task definition and methods}
\label{sec:discrete}

We pose the task of \emph{vocal accompaniment} as a conditional generative modeling problem, where the goal is to model a distribution  
${P(\inswav \mid \voxwav)}$ 
over appropriate instrumental waveforms~$\inswav$ for vocal waveforms~$\voxwav$, where both waveforms are monaural, $T$ seconds in length, and sampled at some rate $f_s$. 
Formally, both $\voxwav$ and $\inswav$ are vectors in $\mathbb{R}^{f_sT}$. 
Given vocals $\sample{\voxwav}$ from a user, 
we can generate music containing those vocals by sampling 
${\sample{\inswav} \sim P(\inswav \mid \voxwav = \sample{\voxwav})}$
and linearly mixing the outputs (e.g. 
${\sample{\voxwav} + \sample{\inswav}}$).

\subsection{Modeling proxy distributions of audio codes}
\label{sec:proxy}

Directly modeling distributions over waveforms is empirically challenging due to the high sampling rates needed for representing audio. 
Adopting a practice first proposed by~\citet{van2017neural} for unconditional generative modeling of audio, 
in this work we instead model discrete audio \emph{codes}. 
Such an approach requires a \emph{discrete codec}, a pair of functions 
${\enc : \mathbb{R}^{f_sT} \to \mathbb{V}^{f_kT}}$ and 
${\dec : \mathbb{V}^{f_kT} \to \mathbb{R}^{f_sT}}$, 
where 
$\mathbb{V}$ is some finite vocabulary, 
$f_k$ is the audio code rate,
and 
$\dec$ approximates $\enc^{-1}$. 

Using these two functions, we can instead model a ``proxy'' distribution over codes produced by $\enc$, and approximate the distribution over waveforms 
by 
leveraging $\dec$. 
Specifically, for an unconditional setting where the goal is to model a distribution over waveforms $\somewav$, we might instead model $P(\somecodes)$ as a proxy, where 
${\somecodes = \enc(\somewav)}$. 
To sample audio from this distribution, we first sample 
${\sample{\somecodes} \sim P(\somecodes)}$ and output 
${\dec(\sample{\somecodes})}$.
Because $f_k \ll f_s$ in practice, modeling this proxy is empirically tractable. 

\begin{figure*}[t]
\begin{center}
\centerline{\includegraphics[width=0.98\textwidth]{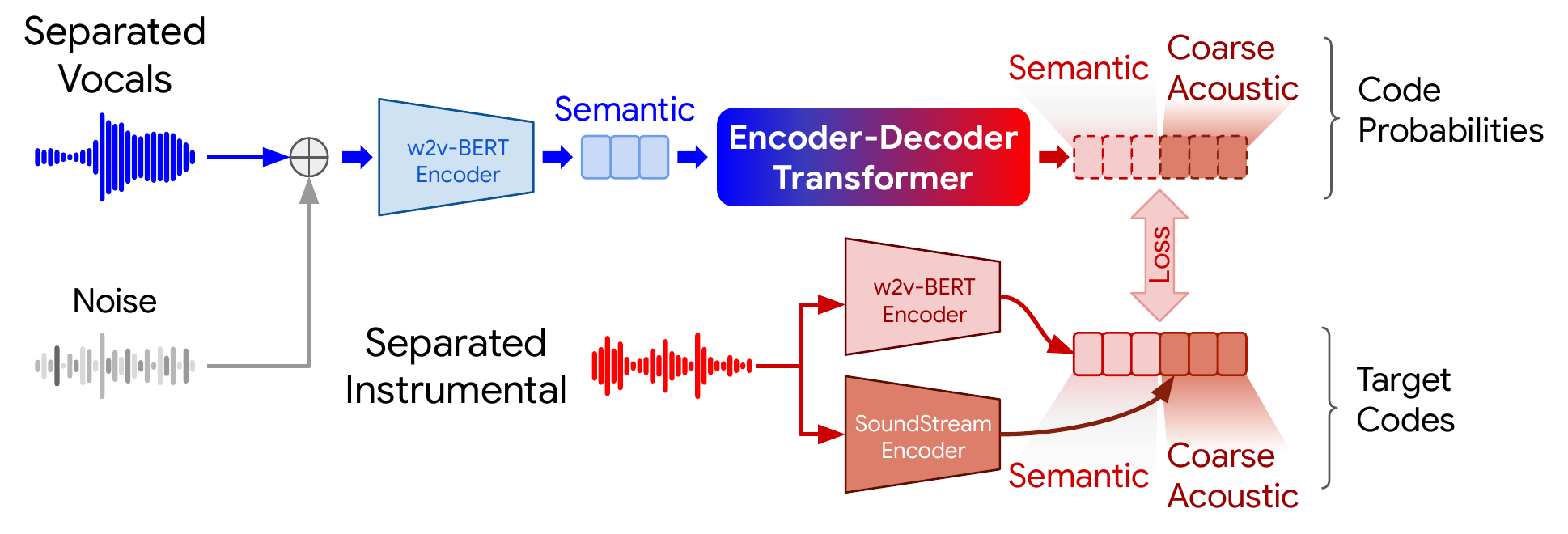}}
\caption{
We adapt \audiolm{}~\citep{borsos2022audiolm} to be suitable for training conditional ``audio-to-audio'' generative models of instrumentals given vocals. 
During training, we use source-separated vocals and instrumentals as inputs and targets respectively. 
We add white noise to the input to conceal residual artifacts of the instrumental present in the source-separated vocals. 
For the targets, we reuse the discrete featurization scheme from \audiolm{}, 
extracting \emph{semantic} codes from a pre-trained \wtvbert{} model~\citep{chung2021w2v} and \emph{coarse acoustic} codes from a pre-trained \soundstream{} codec~\citep{zeghidour2021soundstream}. 
We experiment with several featurizations of the input---our best model uses only semantic codes of the noisy vocals. 
We train T5~\citep{raffel2020exploring}, an encoder-decoder Transformer~\citep{vaswani2017attention}, 
to predict target codes given input codes. 
During inference (not shown), we use this model to generate coarse acoustic codes, 
then use a separately-trained model to generate fine codes given coarse, 
and finally decode both with \soundstream{}.
}
\label{fig:overview}
\end{center}
\vspace{-2em}
\end{figure*}

\subsection{AudioLM preliminaries}
\label{sec:audiolm_preliminaries}

In this work, we base our method on \audiolm~\citep{borsos2022audiolm}, 
a state-of-the-art unconditional generative model of audio, which uses a factorized approach to model a proxy distribution over multiple types of audio codes. 

\paragraph{Acoustic codes} The discrete codec in \audiolm{} is a pre-trained \soundstream{}~\cite{zeghidour2021soundstream} model with an encoder ${\acoenc : \mathbb{R}^{f_sT} \to \mathbb{V}^{50T \times 12}}$ 
and a corresponding decoder $\acodec$, where ${f_s = 16}$kHz and ${|\mathbb{V}| = 1024}$. 
The outputs from $\acoenc$ are referred to as \emph{acoustic codes}. 
\soundstream{} uses a residual vector quantization scheme producing $12$-dimensional vectors of acoustic codes at a rate of \units{50}{Hz}---the first $4$ dimensions and the remaining $8$ are the \emph{coarse} and \emph{fine} acoustic codes, respectively. 
\audiolm{} ``flattens''
these coarse and fine codes, 
inducing rates of $200$ and $400$ codes per second of audio respectively. 
We denote the flattened coarse acoustic codes for waveforms $\somewav$ as 
${\acocoarse{\somewav}}$ 
and flattened fine acoustic codes as 
${\acofine{\somewav}}$, 
which can be trivially recombined as ${\acoenc(\somewav)}$.
Note that, because of its residual quantization scheme, \soundstream{} can decode audio from coarse codes alone, though decoding from coarse \emph{and} fine codes yields higher audio fidelity.

\paragraph{Semantic codes} Past music audio generative models such as Jukebox~\citep{wu2022jukedrummer} have directly modeled representations similar to acoustic codes with larger models. 
\audiolm{} instead proposes to use smaller models to model a joint distribution over acoustic codes and low-rate \emph{semantic codes}, which originate from a semantic encoder ${\semenc : \mathbb{R}^{f_sT} \to \mathbb{S}^{25T}}$ operating at \units{25}{Hz} and trained with a self-supervised loss instead of an acoustic reconstruction loss. 
The semantic encoder consists of intermediate representations from a pre-trained \wtvbert{} model~\citep{chung2021w2v} that have been quantized with $k$-means, where ${k = |\mathbb{S}| = 1024}$. 

\setlength{\abovedisplayskip}{8pt}
\setlength{\belowdisplayskip}{8pt}

\audiolm{} is a cascade of three models that generate increasingly high-rate codes. 
Specifically, \audiolm{} factorizes the joint distribution over semantic and acoustic codes as: 
\begin{align*}
P(&\semenc(\somewav), \acoenc(\somewav)) = \\
&P(\acofine{\somewav} \mid \acocoarse{\somewav}) \\
\cdot 
&P(\acocoarse{\somewav} \mid \semenc(\somewav)) \\
\cdot 
&P(\semenc(\somewav)), 
\end{align*}
with the assumption 
${\acofine{\somewav} \CI \semenc(\somewav) \mid \acocoarse{\somewav}}$, 
i.e.~that semantic codes are unhelpful for generating fine acoustic codes given coarse ones. 
Each model is an autoregressive, decoder-only Transformer~\citep{vaswani2017attention} ``language model'' that is trained separately, and conditioning is achieved by simply concatenating the two sequences.

To generate audio with \audiolm, the three generative models are first sampled in series. 
Then, the generated semantic codes are discarded as a byproduct, 
and the generated acoustic codes are fed to $\dec$, 
producing a waveform. 

\subsection{Adapting \audiolm{} for accompaniment}
\label{sec:audiotoaudiolm}

Instead of directly modeling distributions over instrumental waveforms~$\inswav$ for vocal waveforms~$\voxwav$, 
in this work we adapt \audiolm{} to model proxy distributions over instrumental codes given vocal codes. 
Specifically, we model 
${P(\semenc(\inswav), \acoenc(\inswav) \mid \feats(\voxwav))}$, 
where $\semenc$ and $\acoenc$ are from \audiolm{}, 
and 
${\feats : \mathbb{R}^{f_sT} \to \mathbb{X}^{f_xT}}$ is a function to extract discrete features from vocals $\voxwav$. 
We adopt a three-stage factorization reminiscent of that of \audiolm:
\begin{align*}
P(&\semenc(\inswav), \acoenc(\inswav) \mid \feats(\voxwav)) = \\
&P(\acofine{\inswav} \mid \acocoarse{\inswav}) \\
\cdot &P(\acocoarse{\inswav} \mid \semenc(\inswav), \feats(\voxwav)) \\
\cdot &P(\semenc(\inswav) \mid \feats(\voxwav)),
\end{align*}
with an analogous independence assumption, i.e.~that fine instrumental codes can be reconstructed from coarse instrumental codes alone.\footnote{Formally, ${\acofine{\inswav} \CI \semenc(\inswav), \feats(\voxwav) \mid \acocoarse{\inswav}}$.}

For \emph{Stage 3}---the model that predicts fine acoustic codes from coarse---we reuse the same approach from \audiolm{}, i.e.,~a decoder-only Transformer language model. 
For convenience, we combine the first two stages into a single ``sequence-to-sequence'' language model that predicts instrumental semantic and coarse acoustic codes given vocal features (\Cref{fig:overview}). 
Specifically, we construct ``input'' sequences ${\voxcodes = \feats(\voxwav)}$, 
and ``target'' sequences 
${\inscodes =[\inscodes_1, \ldots, \inscodes_{f_kT}] \in (\mathbb{S} \cup \mathbb{V})^{f_kT}}$ by concatenating 
${[\semenc(\inswav) ; \acocoarse{\inswav}]}$, where $f_k = 225$ codes per second. 
Then, we model
\begin{align*}
P_{\theta,\phi}(\inscodes \mid \voxcodes) &= 
\prod_{t=1}^{Tf_k} P_{\theta} (\inscodes_{t} \mid \inscodes_{<t}, E_{\phi}(\voxcodes)), \text{where} \\
P_{\theta} (\inscodes_{t} \mid \inscodes_{<t}, E_{\phi}(\voxcodes)) &= 
\texttt{SoftMax}(D_{\theta}(\inscodes_{<t}, E_{\phi}(\voxcodes))).
\end{align*}
For our sequence encoder ($E_{\phi}$) and decoder ($D_{\theta}$), 
we adopt the encoder-decoder architecture from T5~\citep{raffel2020exploring}, 
which uses Transformers~\citep{vaswani2017attention}. 

To use this trained model in our application given vocal input $\sample{\voxwav}$, we first sample instrumental semantic and coarse acoustic codes
${\sample{\inscodes} \sim P_{\theta,\phi}(\inscodes \mid \voxcodes = \feats(\sample{\voxwav}))}$. 
Then, we drop the semantic codes and use Stage 3 to generate fine acoustic codes given the coarse codes. 
Finally, we combine the generated coarse and fine acoustic codes as 
${\texttt{Combined}(\sample{\inscodes})}$ and output 
${\sample{\voxwav} + \dec(\texttt{Combined}(\sample{\inscodes}))}$. 
Note that we specifically combine the \emph{original} vocal input $\sample{\voxwav}$ with the \emph{decoded} instrumental ${\texttt{Combined}(\sample{\inscodes})}$, 
as opposed to a version of the vocal input that has been round tripped through the codec. 

\subsection{Featurizing input vocals}
\label{sec:feats}

Here we describe the function ${\feats : \mathbb{R}^{f_sT} \to \mathbb{X}^{f_xT}}$
that we use to extract discrete features from vocal inputs. 
For symmetry with our target sequences, we adopt a default that naively reuses the same featurization as that of the instrumental targets:
${\feats_{\text{Def.}}(\voxwav) = [\semenc(\voxwav) ; \acocoarse{\voxwav}]}$, i.e.~$f_x = 225$Hz and 
${\mathbb{X} = \mathbb{S} \cup \mathbb{V}}$.

\paragraph{Adding noise.}
We propose to add white noise to vocal inputs during both training and inference. 
The primary purpose of adding noise is to conceal barely-audible artifacts of the original instrumental that remain in source-separated vocals (see our sound examples page in footnote of first page to listen to these artifacts). 
Specifically, we define a function 
${\texttt{Noise} : \voxwav \mapsto \voxwav + \noisewav}$, where 
${\noisewav \sim \mathcal{N}(0,\sigma^2)^{Tf_s}}$ and $\sigma$~is a hyperparameter controlling the amplitude of the noise. Using this, we define 
${\feats(\voxwav) = \feats_{\text{Def.}}(\texttt{Noise}(\voxwav))}$ as our feature extraction function.

In preliminary experiments training on source-separated data, models exhibited poor generalization. 
Specifically, 
when fed source-separated vocals, 
models would output instrumentals suspiciously reminiscent of the originals, 
and would output mostly silence when fed isolated vocals. 
We suspected that we had inadvertently trained models to reconstruct the original instrumental from its artifacts. 
By adding noise at an amplitude above that of the instrumental artifacts but below that of the vocals, 
we hypothesize that we can discourage the model from relying on these artifacts during training, hopefully improving generalization.

In Section~\ref{sec:features} we explore a range of featurizations including additive noise and different combinations of semantic and acoustic codes. 
We leverage the hierarchy of \audiolm{} and flexibility of conditioning to identify featurizations that are robust to the artifacts of source separation.

\section{Experiments and results}

Here we describe our experimental protocol for training and evaluating \sys{}.

\subsection{Datasets}
\label{sec:datasets}

\paragraph{Training (source-separated).} The training set for \sys{} 
is comprised of
$1$ million audio-only sources resulting in $46$k hours of music. 
We preprocess this dataset by resampling each source mix from its original sampling rate to \units{44.1}{kHz} 
(the majority do not require resampling) 
and averaging stereo mixes to mono. 
From here, our preprocessing differs for \emph{pre-training} \soundstream{} and \wtvbert{} vs. \emph{training} \sys{}. 
For \soundstream{} and \wtvbert{}, we resample mixes to \units{16}{kHz}, extract non-overlapping \units{30}{s} clips, and pre-train on all clips. 
For training \sys{}, we preprocess further by extracting non-overlapping \units{10}{s} clips from each mix and input resultant clips to the \mdxnet{}~\citep{kim2021mdx} source separation algorithm (which natively operates at \units{44.1}{kHz}) to extract vocals. 
We subtract the source-separated vocals from the original mix to yield a corresponding source-separated instrumental. 
Finally, we resample vocal and instrumental clips to \units{16}{kHz}, the sampling rate of \soundstream{} and \wtvbert{}. 

\paragraph{Evaluation (isolated).} To evaluate model performance on isolated vocals, we adopt the MUSDB18 dataset~\citep{rafii2017musdb18}, which contains $10$ hours of professional, studio-isolated vocal and instrumental ``stems'' for $150$ source mixes. 
We identify non-overlapping $10$s clips from each mix where the vocals are present (defined as peak RMS amplitude of vocals above \units{-25}{dB} relative to line level), 
resulting in $1232$ clips for the training set and $778$ clips for the test set.
We also run each $10$s clip from the original mixture through \mdxnet{} to extract source-separated vocals, allowing us to directly compare model performance on source-separated and isolated vocals for the same musical material. 
Because we do not train on MUSDB18 but use its training set for hyperparameter comparisons, we henceforth refer to it as our ``dev'' set.

\subsection{Filtering training data}
\label{sec:filter}

In an effort to increase the relevance of our training data to the task of interest, 
we propose a filtering strategy that filters out clips where 
(1)~the instrumental is silent, or
(2)~the vocals are substantially louder than the instrumental. 
Specifically, we filter out clips where the peak RMS amplitude of the instrumental is below \units{-25}{dB} (relative to line level), 
or clips where the peak RMS amplitude of the vocals is at least \units{5}{dB} louder than that of the instrumental. 
The goal of filtering is to bias the system towards always outputting \emph{some} instrumental for all inputs, 
which is more useful than a system which respects the unfiltered training data distribution by occasionally outputting quiet or silent instrumentals.

\subsection{Evaluation}

As our primary quantitative evaluation metric, 
we adopt the Fr\'echet Audio Distance~(\fad) metric proposed by~\citet{kilgour2019fad}, which is shown to correlate more closely with human perception of audio quality than other conventional automated metrics. 
\fad{} is the audio analogue to the FID metric~\citep{heusel2017fid} commonly used for benchmarking image generation models. 
Specifically, \fad{} is the Fr\'echet distance~\citep{frechet1957distance} between two multivariate Gaussians computed over embeddings from the VGGish audio classifier~\citep{hershey2017cnn} of the reference and generated audio respectively. 

In our setting, we compute \fad{} on MUSDB18 (preprocessed as described in~\Cref{sec:datasets}), 
using the ground truth mixes as the reference audio. 
To measure the generalization of models trained as part of this work, 
we separately feed in isolated and source-separated vocals as inputs, 
combining the output instrumental with isolated vocals in both scenarios to produce mixes for computing \fad{} against the references. 
The scenario where we feed in source-separated vocals is not representative of inference scenarios we expect to encounter in-the-wild---we
specifically refer to computing \fad{} on isolated vocal inputs as \fadi{}, and on source-separated vocal inputs as \fads{}.
We refer to their difference ($\Delta$) as the \emph{generalization gap}. 

We also compute the negative log-likelihood~(\nll) of our models over the coarse acoustic codes from the isolated instrumentals given source-separated and isolated vocal inputs. 
Note that, in practice, we found \nll{} to be poorly correlated with our own perception of quality---all models improved more-or-less monotonically on \nll{} during training despite some models collapsing both subjectively and on \fadi{} (see~\Cref{fig:nllbad} for an example).
Nonetheless, we report \nll{} in our appendix (\Cref{tab:quant_verbose}) to give a more complete picture of our experiments.

\begin{figure}[t]
\begin{center}
\centerline{\includegraphics[width=0.98\columnwidth]{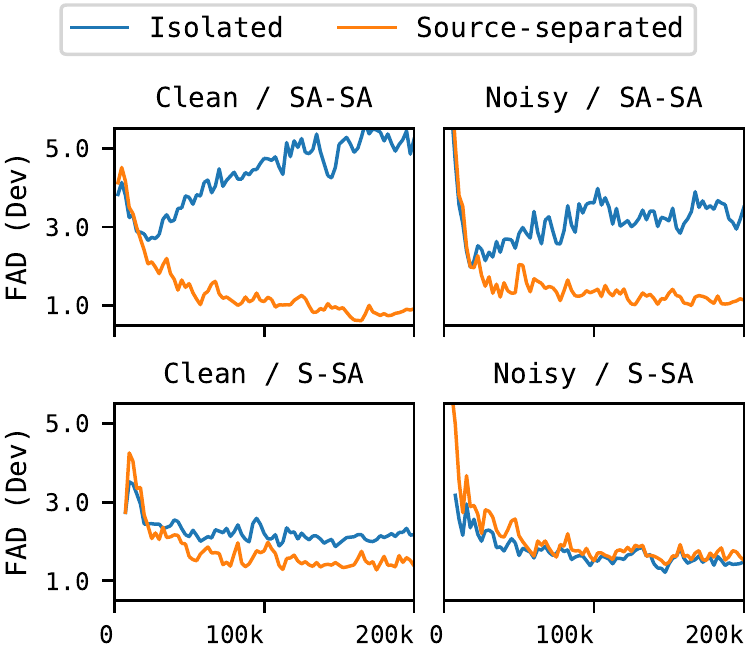}}
\caption{
We experiment with different featurizations of input vocals to improve our system's ability to generalize from source-separated vocals seen during training to the isolated vocals we anticipate receiving from users. 
We found that the combination of
(1)~\texttt{Noisy}: adding white noise to vocal inputs to mask source separation artifacts, and
(2)~\texttt{S-SA}: removing acoustic codes of vocals from the default \audiolm{} featurization (\texttt{SA-SA}), 
resulted in strong generalization. 
Namely, training with this featurization results in nearly identical FAD when evaluating on MUSDB18-dev with isolated and source-separated vocals (bottom right).
}
\label{fig:generalization}
\end{center}
\vspace{-2em}
\end{figure}

\subsection{Modeling hyperparameters}
\label{sec:model}
\looseness=-1
Here we outline the default modeling configuration for our experiments. 
We adopt the default architecture and training hyperparameters from the \texttt{t5.1.1.base} configuration\footnote{\texttt{t5.1.1.base}: \url{https://bit.ly/3Hy91wp}} implemented in \texttt{t5x} \citep{roberts2022t5x} with the following changes:
(1)~we increase dropout from $0$ to $0.1$, 
(2)~we replace the relative positional embeddings with the fixed positional encodings from vanilla Transformer~\citep{vaswani2017attention}, which are more computationally efficient for longer sequences and performed marginally better on our task in preliminary experiments. 
We train models for $200$k steps (batch size $512$) on \units{10}{s} clips inducing maximum sequence lengths of $2250$ for both inputs and targets. 
We regularize models with early stopping based on \fadi{} computed on MUSDB18-dev.
As is the case in \audiolm{}, here we use disjoint vocabulary subsets to represent each of the $4$ coarse acoustic code dimensions, inducing a vocabulary size of $5122$: $1024$ semantic, $4096$ acoustic, $2$ sentinel tokens (start-of-sequence and padding---the latter is unused). 
Each sequence index is uniquely tied to a specific subset of this vocabulary---during sampling, we manually enforce this behavior by truncating invalid subsets from the distribution. 
We decode with temperature sampling using a temperature of $0.85$.

\subsection{Audio featurization experiments}
\label{sec:features}

We explore several different audio featurization configurations for both the input vocals and the target instrumentals. 
For the input vocals, we use the \texttt{Feats} function as described in~\Cref{sec:feats}, which concatenates semantic and coarse acoustic codes for vocals with optional additive noise. 
We experiment with two noise conditions: 
\texttt{Clean}, where $\sigma = 0$ (no noise), and \texttt{Noisy}, where $\sigma=0.01$, i.e.,~vocals are combined with noise attenuated by $40$dB (determined through subjective listening to be louder than most artifacts but quieter than the vocals). 
As a default for the target, we also concatenate semantic and coarse acoustic codes for the instrumental with no noise, as outlined in~\Cref{sec:audiotoaudiolm}. 
We refer to this default featurization as \texttt{SA-SA}, i.e.,~input is semantic/acoustic and target is semantic/acoustic.
We also experiment with dropping coarse acoustic codes from the input conditioning (\texttt{S-SA}).

In~\Cref{fig:generalization}, we show that different input vocal featurizations produce different generalization properties. 
Both the \texttt{Noisy} and the \texttt{S-SA} configurations independently seem to improve generalization, i.e.,~they reduce the generalization gap of the model when fed source-separated (as in training) vs. isolated vocals (as is more realistic at inference time). 
Together, both featurizations essentially eliminate the generalization gap, leading to stable training dynamics and reducing the importance of early stopping.

\begin{table}[t]
\vskip 0.15in
\centering
\raisebox{-.83in}{\rotatebox{90}{\raisebox{.02in}{$\underset{\overbrace{\hspace{1.05in}}{}}{\texttt{Noisy}}$}\hspace{.05in}$\underset{\overbrace{\hspace{.35in}}{}}{\texttt{Clean}}$}}
\begin{tabular}{lcccc}
\toprule
\tabonerow{Method}{\text{Steps}}{\fadi}{\fads}{$\Delta$}{\nlli}{\nlls}
\midrule
\tabonerow{\expname{SA-SA}}{$23$k}{$3.01$}{$1.61$}{$1.39$}{$3.84$}{-}
\tabonerow{\expname{S-SA}}{$148$k}{$2.31$}{$\mathbf{1.14}$}{$1.17$}{$3.72$}{$3.71$}
\midrule
\tabonerow{\expname{SA-SA}}{$18$k}{$2.01$}{$1.64$}{$0.37$}{$3.92$}{$3.89$}
\tabonerow{\expname{A-A}}{$18$k}{$3.41$}{$3.30$}{$\mathbf{0.11}$}{$4.03$}{$3.99$}
\tabonerow{\expname{SA-A}}{$43$k}{$2.81$}{$1.87$}{$0.95$}{$3.94$}{$3.89$}
\tabonerow{\expname{A-SA}}{$48$k}{$2.01$}{$1.65$}{$0.36$}{$3.85$}{$3.79$}
\tabonerow{\expname{S-SA}}{$148$k}{$\mathbf{1.36}$}{$1.17$}{$0.19$}{$\mathbf{3.71}$}{$3.70$}
\midrule
\tabonerow{\expname{S-SA-XL}}{$150$k}{$\mathbf{1.28}$}{$\mathbf{0.96}$}{$0.32$}{$\mathbf{3.47}$}{$\mathbf{3.47}$}
\bottomrule
\end{tabular}
\caption{
Quantitative evaluation for our experiments. \fad{} is the Fr\'echet Audio Distance on MUSDB18-test between ground truth mixes and mixes of ground truth vocals and instrumentals generated by models using either isolated vocals (\fadi{}) or source-separated vocals (\fads{}) as inputs. Since \fadi{} corresponds better to the anticipated usage, and \fads{} is closer to how the model is trained, a generalization gap ($\Delta$) exists between the two. Adding noise to the input vocals (\texttt{Noisy}), 
removing the vocal acoustic codes (\expname{S-SA}), 
and increasing model size (\expname{XL}) 
all help to improve generation to isolated vocal inputs (\fadi{}).
Full details of featurizations can be found in Section~\ref{sec:features}. For all experiments, \fad{} on MUSDB18-dev is used as an early stopping metric.}
\label{tab:quant}
\vskip -0.1in
\end{table}

We also experiment with additional featurizations for the input and target to better understand the effects of each feature. 
We drop the semantic codes from the 
input~(\texttt{A-SA}), 
target~(\texttt{SA-A}), and 
both the input and target~(\texttt{A-A})---we cannot drop the acoustic codes from the target as they are necessary for synthesis. 
We run these additional experiments only on the \texttt{Noisy} condition due to the poor performance of both models trained under the \texttt{Clean} condition and high computational cost of experiments. 
Results for all models are in~\Cref{tab:quant}. 

Overall, our best model (\texttt{Noisy}~/~\texttt{S-SA}) adds noise to vocal inputs, 
uses only semantic codes for vocals as conditioning info, 
and uses both semantic and coarse acoustic codes for target instrumentals. 
This model improves performance on \fadi{} by $55$\% relative to a naive adaptation of \audiolm{} (\texttt{Clean}~/~\texttt{SA-SA}). 
Removing semantic codes from the instrumental hurt performance (relative to \texttt{SA-SA}) in both experiments (\texttt{SA-A} and \texttt{A-A}). 
Only using acoustic codes for the vocals (\texttt{A-SA}) resulted in similar performance to using both semantic and acoustic codes. 
All of the models that included vocal acoustic codes as conditioning info diverged in performance on isolated vocals after fewer than $50$k steps.

\paragraph{Scaling.}
We take our best model configuration (\texttt{Noisy}~/~\texttt{S-SA}) and scale up from the \texttt{t5.1.1.base} configuration ($250$M parameters) to \texttt{t5.1.1.xl} ($3$B parameters) keeping all other hyperparameters fixed, which we refer to as \texttt{S-SA-XL}. 
Perhaps unsurprisingly, we find that scaling up does improve quantitative performance compared to \texttt{S-SA} at smaller scale. 
We henceforth refer to our best models as \texttt{SingSong-Base} and \texttt{SingSong-XL}.

\section{Listening study}

We conduct a listening study to measure qualitative performance of our two best models against the ground truth and baselines. 
For our study, listeners are presented a pair of $10$s vocal-instrumental mixtures, where the vocals are identical between the two mixtures and come from MUSDB18-test, and the instrumentals come from different sources (ground truth, our models, or baselines). 
Listeners are asked to indicate in which of the two mixtures do the instrumental accompaniments seem more musically compatible with the vocals. 
Listeners are explicitly discouraged from paying attention to the audio fidelity of the instrumental.

\subsection{Baselines}
\label{sec:baselines}
We compare to two retrieval-based baselines: 
one that retrieves an instrumental clip uniformly at random from the MUSDB18-dev set (\texttt{Random}), and 
one that uses musical features of the ground truth mixture to retrieve an appropriate instrumental from MUSDB18-dev and adapt it to further improve alignment (\texttt{Retrieval}).
The latter is similar in design to Songsmith~\citep{simon2008mysong} except that it retrieves instrumental audio instead of instrumental MIDI---we do not directly compare to Songsmith as it does not have an API allowing for input of pre-recorded vocals. 

To construct \texttt{Retrieval}, we first run each ground truth mixture in MUSDB18-dev through a musical key detection system~\cite{korzeniowski2018genreagnostic} that estimates probabilities of the mixture being in each of $24$ keys ($12$ major, $12$ minor). 
We also estimate the tempo of each ground truth instrumental using a tempo detection system~\cite{bock2015accurate}---both key and tempo features were computed using the  \texttt{madmom} library~\cite{bock2016madmom}. 
Then, for an input vocal query from MUSDB18-test, we use the same key detection system to estimate key probabilities of the vocals, and we detect tempo from the ground truth instrumental associated with that vocal input. 
We select the instrumental track from MUSDB18-dev that has the lowest Euclidean distance in key probability space, then time stretch the instrumental such that its new tempo matches the estimated tempo of the input. 
Note that, because the query tempo is estimated from associated instrumentals instead of the input vocals, this baseline receives extra information that would not be available in realistic settings---we found tempo estimation directly from input vocals to be too unreliable.

\subsection{Results}

\begin{figure}[t]
\begin{center}
\centerline{\includegraphics[width=0.98\columnwidth]{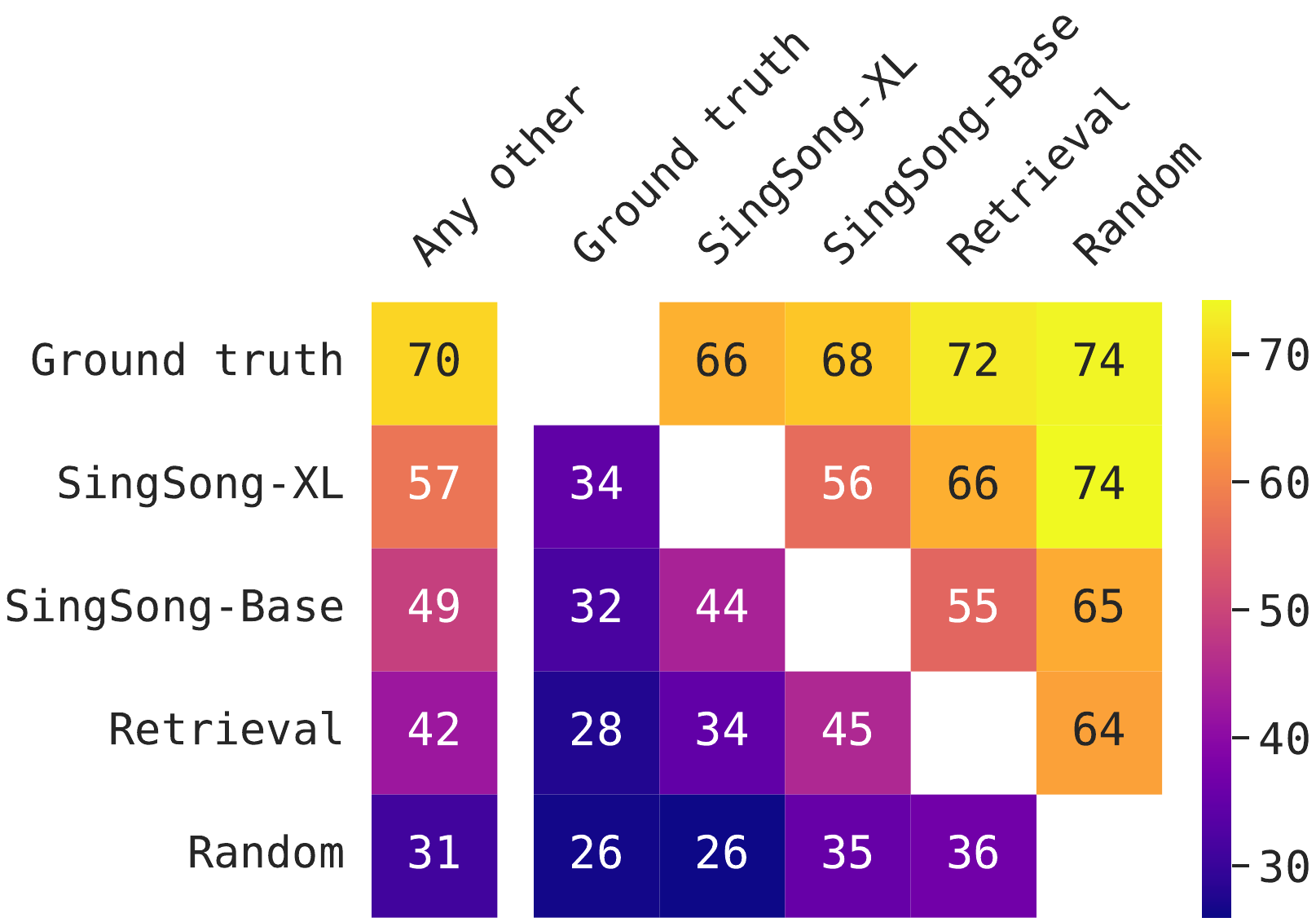}}
\caption{Results from our qualitative study where listeners judged pairs of vocal-instrumental mixtures where vocals are fixed and instrumentals come from two different sources (ground truth, \sys, or baselines). Each row indicates the \% of times listeners preferred instrumentals from that system compared to those from any other system (first column, $N = 800$) and each system individually (other columns, $N = 200$).}
\label{fig:listening}
\end{center}
\vspace{-2em}
\end{figure}

Results for all systems appear in~\Cref{fig:listening}. 
When comparing our best system (\texttt{SingSong-XL}) to our strongest baseline (\texttt{Retrieval}), 
listeners preferred instrumentals from our best system in $66$\% of cases. 
A Wilcoxon signed-rank test of these pairwise judgements indicates that listeners preferred instrumentals from \texttt{SingSong-Base} and \texttt{XL} significantly more often than those of the strongest baseline (${p=0.01}$ and ${p=2 \times 10^{-6}}$, respectively). 
Results also indicate significant qualitative improvements from scaling---listeners preferred instrumentals from \texttt{SingSong-XL} in $56$\% of comparisons to \texttt{SingSong-Base} (${p=0.01}$ with the same statistical test). 
Moreover, listeners preferred instrumentals from our best system in $57$\% of cases against any other source including the ground truth, compared to only $42$\% for our strongest baseline. 
See our sound examples (footnote of first page) to listen to comparisons.

\section{Discussion}

Overall, \sys{} produces instrumental accompaniments with strong qualitative performance relative to a strong baseline which receives extra information. 
Subjectively speaking, 
\sys{} outputs instrumentals that often have clear harmonic and temporal correspondence to the input vocals. 
One area for improvement is that generated instrumentals tend have weaker (both in volume and coherence) harmonic elements compared to percussive elements. 
While removing the vocal acoustic codes was important for generalization, we suspect that only using vocal semantic codes as conditioning may provide insufficient harmonic context for instrumental generation---we hope to explore other avenues for input featurization in future work. 
Another issue is that the sampling rate of the \audiolm{} acoustic codec (\units{16}{kHz}) is well below conventional sampling rates for high-fidelity music, imposing a low ceiling for our system's fidelity---in future work, we hope to increase the sampling rate of the codec. 

\paragraph{Longer examples.}
On our sound examples page, we also share examples of our best model (\texttt{SingSong-XL}) operating on longer ($30$s) vocal clips. 
To achieve these results, we apply our model on $10$s windows of vocal inputs in a sliding fashion with $50$\% overlap, 
i.e.,~we first generate $10$s of instrumental conditioned on the first $10$s of vocals, then iteratively generate $5$s chunks of instrumental conditioned on $10$s windows of vocals while prompting the decoder with the preceding $5$s of generated instrumental, both at offset $5i$ seconds ($i=1$ on the first iteration). 
Subjectively speaking, 
despite being conditioned only on local context ($10$s of vocals and the previous $5$s of generated instrumental), 
this approach achieves reasonable global coherence on $30$s vocal inputs, with style and instrumentation remaining fairly consistent across windows. 

\paragraph{Real-world inputs.}
On our sound examples page, we also include results of our system operating on clips from the Vocadito dataset~\citep{bittner2021vocadito}, which contains vocals from novice singers recorded on a variety of commodity microphones and mobile devices. 
While we do not evaluate on this dataset as it lacks ground truth instrumentals, 
it may be more representative than MUSDB18 of inputs we expect to see from users of \sys{} in the real world. 
Subjectively speaking, results on Vocadito are promising though perhaps not as strong as results on MUSDB18---in future work, we hope to explore adjusting user \emph{inputs} to our system by having users sing to a clicktrack as in~\citet{simon2008mysong}, or applying pitch correction.

\paragraph{Memorization analysis.}

We conduct a preliminary analysis to measure memorization in \texttt{SingSong-XL} and find that, in general, training data cannot be extracted verbatim from our model. 
We use a protocol similar to the one described in~\citet{carlini2022quantifying}. 
Specifically, we pass in vocal inputs from the training data, prompt our decoder with the first $k$ semantic codes of the ground truth instrumental, and report the number of exact matches against the ground truth when decoding the remaining ${250-k}$ instrumental semantic codes with greedy sampling. 
For each of ${k \in \{0,25,50,125\}}$, we find \emph{no} exact matches in $8$k trials. 
For ${k=225}$, we find only $4$ exact matches ($0.05$\%).

\paragraph{Future work.}

In addition to the directions we have already mentioned, promising avenues for future work center around further exploring the relationship between source separation and audio-to-audio accompaniment generation. 
Firstly, musical source separation can already extract other sources in addition to vocals and instrumentals, such as drums, bass, and piano. 
In future work, we hope to train accompaniment models to be able to generate any source(s) from any other source(s). 
Source separation and accompaniment generation may also have a symbiotic relationship. 
Improvements in source separation quality (fewer artifacts) may reduce the generalization gap of accompaniment systems, and increases in source separation flexibility (more instrument categories) may result in new accompaniment capabilities (e.g.,~bagpipe accompaniment). 
Similarly, improvements in accompaniment may be useful for generating synthetic but realistic training data for source separation, e.g.~one could apply \sys{} to a large corpus of isolated singing to produce aligned (mix, vocal) pairs.

\paragraph{Ethical considerations.}

While music generation technology presents an opportunity to increase access to creative participation in music, it also presents profound risks---music generation technology may fundamentally reshape music culture and redefine the economic relationships between stakeholders~\citep{detweiler2022redefining}. 
More specific to this work, singing voice evokes perhaps the strongest associations with individual identity of any musical instrument. 
By building a music generation system that 
(1)~requires initiative from users (singing) to produce music, and (2)~preserves the identity of individuals in the output, 
we hope that our system may increase creative participation in music while avoiding the pitfalls of systems that generate music from scratch or mimic the identities of existing singers. 
One risk of our current system is that users have no control over the genre and style of the output instrumental---our system may already be biased
towards 
correlations between cultural identities of users and musical genres associated more prominently with certain cultural identities. 
In future work, we hope to condition our system on genre, style, or text descriptions as in~\citet{agostinelli2023musiclm}, mitigating this risk by offering users explicit control over these high-level attributes.



\bibliography{main}

\begin{thebibliography}{38}
\providecommand{\natexlab}[1]{#1}
\providecommand{\url}[1]{\texttt{#1}}
\expandafter\ifx\csname urlstyle\endcsname\relax
  \providecommand{\doi}[1]{doi: #1}\else
  \providecommand{\doi}{doi: \begingroup \urlstyle{rm}\Url}\fi

\bibitem[Agostinelli et~al.(2023)Agostinelli, Denk, Borsos, Engel, Verzetti,
  Caillon, Huang, Jansen, Roberts, Tagliasacchi, Sharifi, Zeghidour, and
  Frank]{agostinelli2023musiclm}
Agostinelli, A., Denk, T.~I., Borsos, Z., Engel, J., Verzetti, M., Caillon, A.,
  Huang, Q., Jansen, A., Roberts, A., Tagliasacchi, M., Sharifi, M., Zeghidour,
  N., and Frank, C.
\newblock {MusicLM}: Generating music from text.
\newblock \emph{arXiv:2301.11325}, 2023.

\bibitem[Bittner et~al.(2021)Bittner, Pasalo, Bosch, Meseguer-Brocal, and
  Rubinstein]{bittner2021vocadito}
Bittner, R.~M., Pasalo, K., Bosch, J.~J., Meseguer-Brocal, G., and Rubinstein,
  D.
\newblock {vocadito}: A dataset of solo vocals with $f_0$, note, and lyric
  annotations.
\newblock \emph{arXiv:2110.05580}, 2021.

\bibitem[B{\"o}ck et~al.(2016)B{\"o}ck, Korzeniowski, Schl{\"u}ter, Krebs, and
  Widmer]{bock2016madmom}
B{\"o}ck, S., Korzeniowski, F., Schl{\"u}ter, J., Krebs, F., and Widmer, G.
\newblock \texttt{madmom}: a new {Python} audio and music signal processing
  library.
\newblock In \emph{ACM International Conference on Multimedia}, 2016.

\bibitem[Borsos et~al.(2022)Borsos, Marinier, Vincent, Kharitonov, Pietquin,
  Sharifi, Teboul, Grangier, Tagliasacchi, and Zeghidour]{borsos2022audiolm}
Borsos, Z., Marinier, R., Vincent, D., Kharitonov, E., Pietquin, O., Sharifi,
  M., Teboul, O., Grangier, D., Tagliasacchi, M., and Zeghidour, N.
\newblock {AudioLM}: a language modeling approach to audio generation.
\newblock \emph{arXiv:2209.03143}, 2022.

\bibitem[Böck et~al.(2015)Böck, Krebs, and Widmer]{bock2015accurate}
Böck, S., Krebs, F., and Widmer, G.
\newblock Accurate tempo estimation based on recurrent neural networks and
  resonating comb filters.
\newblock In \emph{ISMIR}, 2015.

\bibitem[Caillon \& Esling(2021)Caillon and Esling]{Caillon2021}
Caillon, A. and Esling, P.
\newblock {RAVE}: A variational autoencoder for fast and high-quality neural
  audio synthesis.
\newblock \emph{arXiv:2111.05011}, 2021.

\bibitem[Carlini et~al.(2022)Carlini, Ippolito, Jagielski, Lee, Tramer, and
  Zhang]{carlini2022quantifying}
Carlini, N., Ippolito, D., Jagielski, M., Lee, K., Tramer, F., and Zhang, C.
\newblock Quantifying memorization across neural language models.
\newblock \emph{arXiv:2202.07646}, 2022.

\bibitem[Chung et~al.(2021)Chung, Zhang, Han, Chiu, Qin, Pang, and
  Wu]{chung2021w2v}
Chung, Y.-A., Zhang, Y., Han, W., Chiu, C.-C., Qin, J., Pang, R., and Wu, Y.
\newblock {w2v-BERT}: Combining contrastive learning and masked language
  modeling for self-supervised speech pre-training.
\newblock In \emph{IEEE Automatic Speech Recognition and Understanding Workshop
  (ASRU)}, 2021.

\bibitem[Detweiler et~al.(2022)Detweiler, Coleman, Diaz, Dom, Donahue, Engel,
  Huang, James, Manilow, McCroskery, Pedersen, et~al.]{detweiler2022redefining}
Detweiler, C., Coleman, B., Diaz, F., Dom, L., Donahue, C., Engel, J., Huang,
  C.-Z.~A., James, L., Manilow, E., McCroskery, A., Pedersen, K., et~al.
\newblock Redefining relationships in music.
\newblock \emph{arXiv:2212.08038}, 2022.

\bibitem[Dhariwal et~al.(2020)Dhariwal, Jun, Payne, Kim, Radford, and
  Sutskever]{dhariwal2020jukebox}
Dhariwal, P., Jun, H., Payne, C., Kim, J.~W., Radford, A., and Sutskever, I.
\newblock Jukebox: A generative model for music.
\newblock \emph{arXiv:2005.00341}, 2020.

\bibitem[Dieleman et~al.(2018)Dieleman, van~den Oord, and
  Simonyan]{dieleman2018challenge}
Dieleman, S., van~den Oord, A., and Simonyan, K.
\newblock The challenge of realistic music generation: modelling raw audio at
  scale.
\newblock In \emph{NeurIPS}, 2018.

\bibitem[Donahue et~al.(2019)Donahue, McAuley, and
  Puckette]{donahue2018adversarial}
Donahue, C., McAuley, J., and Puckette, M.
\newblock Adversarial audio synthesis.
\newblock In \emph{ICLR}, 2019.

\bibitem[Engel et~al.(2019)Engel, Agrawal, Chen, Gulrajani, Donahue, and
  Roberts]{engel2019gansynth}
Engel, J., Agrawal, K.~K., Chen, S., Gulrajani, I., Donahue, C., and Roberts,
  A.
\newblock {GANSynth}: Adversarial neural audio synthesis.
\newblock In \emph{ICLR}, 2019.

\bibitem[Engel et~al.(2020)Engel, Hantrakul, Gu, and Roberts]{engel2020ddsp}
Engel, J., Hantrakul, L., Gu, C., and Roberts, A.
\newblock {DDSP}: Differentiable digital signal processing.
\newblock In \emph{ICLR}, 2020.

\bibitem[Fr{\'e}chet(1957)]{frechet1957distance}
Fr{\'e}chet, M.
\newblock Sur la distance de deux lois de probabilit{\'e}.
\newblock \emph{Comptes Rendus Hebdomadaires des Seances de L Academie des
  Sciences}, 1957.

\bibitem[Goel et~al.(2022)Goel, Gu, Donahue, and R{\'e}]{goel2022sashimi}
Goel, K., Gu, A., Donahue, C., and R{\'e}, C.
\newblock It’s raw! audio generation with state-space models.
\newblock In \emph{ICML}, 2022.

\bibitem[Grachten et~al.(2020)Grachten, Lattner, and
  Deruty]{grachten2020bassnet}
Grachten, M., Lattner, S., and Deruty, E.
\newblock {BassNet}: A variational gated autoencoder for conditional generation
  of bass guitar tracks with learned interactive control.
\newblock \emph{Applied Sciences}, 2020.

\bibitem[Hawthorne et~al.(2018)Hawthorne, Stasyuk, Roberts, Simon, Huang,
  Dieleman, Elsen, Engel, and Eck]{hawthorne2018enabling}
Hawthorne, C., Stasyuk, A., Roberts, A., Simon, I., Huang, C.-Z.~A., Dieleman,
  S., Elsen, E., Engel, J., and Eck, D.
\newblock Enabling factorized piano music modeling and generation with the
  {MAESTRO} dataset.
\newblock In \emph{ICLR}, 2018.

\bibitem[Hawthorne et~al.(2022{\natexlab{a}})Hawthorne, Jaegle, Cangea,
  Borgeaud, Nash, Malinowski, Dieleman, Vinyals, Botvinick, Simon,
  et~al.]{hawthorne2022general}
Hawthorne, C., Jaegle, A., Cangea, C., Borgeaud, S., Nash, C., Malinowski, M.,
  Dieleman, S., Vinyals, O., Botvinick, M., Simon, I., et~al.
\newblock General-purpose, long-context autoregressive modeling with {Perceiver
  AR}.
\newblock In \emph{ICML}, 2022{\natexlab{a}}.

\bibitem[Hawthorne et~al.(2022{\natexlab{b}})Hawthorne, Simon, Roberts,
  Zeghidour, Gardner, Manilow, and Engel]{Hawthorne2022}
Hawthorne, C., Simon, I., Roberts, A., Zeghidour, N., Gardner, J., Manilow, E.,
  and Engel, J.
\newblock Multi-instrument music synthesis with spectrogram diffusion.
\newblock In \emph{ISMIR}, 2022{\natexlab{b}}.

\bibitem[Hershey et~al.(2017)Hershey, Chaudhuri, Ellis, Gemmeke, Jansen, Moore,
  Plakal, Platt, Saurous, Seybold, et~al.]{hershey2017cnn}
Hershey, S., Chaudhuri, S., Ellis, D.~P., Gemmeke, J.~F., Jansen, A., Moore,
  R.~C., Plakal, M., Platt, D., Saurous, R.~A., Seybold, B., et~al.
\newblock {CNN} architectures for large-scale audio classification.
\newblock In \emph{ICASSP}, 2017.

\bibitem[Heusel et~al.(2017)Heusel, Ramsauer, Unterthiner, Nessler, and
  Hochreiter]{heusel2017fid}
Heusel, M., Ramsauer, H., Unterthiner, T., Nessler, B., and Hochreiter, S.
\newblock {GANs} trained by a two time-scale update rule converge to a local
  {Nash} equilibrium.
\newblock \emph{NeurIPS}, 2017.

\bibitem[Kilgour et~al.(2019)Kilgour, Zuluaga, Roblek, and
  Sharifi]{kilgour2019fad}
Kilgour, K., Zuluaga, M., Roblek, D., and Sharifi, M.
\newblock {Fr{\'e}chet Audio Distance}: A reference-free metric for evaluating
  music enhancement algorithms.
\newblock In \emph{INTERSPEECH}, 2019.

\bibitem[Kim et~al.(2021)Kim, Choi, Chung, Lee, and Jung]{kim2021mdx}
Kim, M., Choi, W., Chung, J., Lee, D., and Jung, S.
\newblock {KUIELab-MDX-Net}: A two-stream neural network for music demixing.
\newblock In \emph{Proceedings of the MDX Workshop}, 2021.

\bibitem[Korzeniowski \& Widmer(2018)Korzeniowski and
  Widmer]{korzeniowski2018genreagnostic}
Korzeniowski, F. and Widmer, G.
\newblock Genre-agnostic key classification with convolutional neural networks.
\newblock In \emph{ISMIR}, 2018.

\bibitem[Lattner \& Grachten(2019)Lattner and Grachten]{lattner2019high}
Lattner, S. and Grachten, M.
\newblock High-level control of drum track generation using learned patterns of
  rhythmic interaction.
\newblock In \emph{WASPAA}, 2019.

\bibitem[Mehri et~al.(2017)Mehri, Kumar, Gulrajani, Kumar, Jain, Sotelo,
  Courville, and Bengio]{Mehri2017}
Mehri, S., Kumar, K., Gulrajani, I., Kumar, R., Jain, S., Sotelo, J.,
  Courville, A., and Bengio, Y.
\newblock {SampleRNN}: An unconditional end-to-end neural audio generation
  model.
\newblock ICLR, 2017.

\bibitem[Paiement et~al.(2006)Paiement, Eck, and
  Bengio]{Paiement2006ProbabilisticMH}
Paiement, J.-F., Eck, D., and Bengio, S.
\newblock Probabilistic melodic harmonization.
\newblock In \emph{Canadian Conference on AI}, 2006.

\bibitem[Raffel et~al.(2020)Raffel, Shazeer, Roberts, Lee, Narang, Matena,
  Zhou, Li, Liu, et~al.]{raffel2020exploring}
Raffel, C., Shazeer, N., Roberts, A., Lee, K., Narang, S., Matena, M., Zhou,
  Y., Li, W., Liu, P.~J., et~al.
\newblock Exploring the limits of transfer learning with a unified text-to-text
  transformer.
\newblock \emph{JMLR}, 2020.

\bibitem[Rafii et~al.(2017)Rafii, Liutkus, St{\"o}ter, Mimilakis, and
  Bittner]{rafii2017musdb18}
Rafii, Z., Liutkus, A., St{\"o}ter, F.-R., Mimilakis, S.~I., and Bittner, R.
\newblock {MUSDB18} - a corpus for music separation.
\newblock 2017.

\bibitem[Roberts et~al.(2022)Roberts, Chung, Levskaya, Mishra, Bradbury, Andor,
  Narang, Lester, Gaffney, Mohiuddin, Hawthorne, et~al.]{roberts2022t5x}
Roberts, A., Chung, H.~W., Levskaya, A., Mishra, G., Bradbury, J., Andor, D.,
  Narang, S., Lester, B., Gaffney, C., Mohiuddin, A., Hawthorne, C., et~al.
\newblock Scaling up models and data with $\texttt{t5x}$ and $\texttt{seqio}$.
\newblock \emph{arXiv:2203.17189}, 2022.

\bibitem[Simon et~al.(2008)Simon, Morris, and Basu]{simon2008mysong}
Simon, I., Morris, D., and Basu, S.
\newblock {MySong}: automatic accompaniment generation for vocal melodies.
\newblock In \emph{SIGCHI}, 2008.

\bibitem[van~den Oord et~al.(2016)van~den Oord, Dieleman, Zen, Simonyan,
  Vinyals, Graves, Kalchbrenner, Senior, and Kavukcuoglu]{oord2016wavenet}
van~den Oord, A., Dieleman, S., Zen, H., Simonyan, K., Vinyals, O., Graves, A.,
  Kalchbrenner, N., Senior, A., and Kavukcuoglu, K.
\newblock {WaveNet}: A generative model for raw audio.
\newblock \emph{arXiv:1609.03499}, 2016.

\bibitem[van~den Oord et~al.(2017)van~den Oord, Vinyals, and
  Kavukcuoglu]{van2017neural}
van~den Oord, A., Vinyals, O., and Kavukcuoglu, K.
\newblock Neural discrete representation learning.
\newblock In \emph{NeurIPS}, 2017.

\bibitem[Vaswani et~al.(2017)Vaswani, Shazeer, Parmar, Uszkoreit, Jones, Gomez,
  Kaiser, and Polosukhin]{vaswani2017attention}
Vaswani, A., Shazeer, N., Parmar, N., Uszkoreit, J., Jones, L., Gomez, A.~N.,
  Kaiser, {\L}., and Polosukhin, I.
\newblock Attention is all you need.
\newblock In \emph{NeurIPS}, 2017.

\bibitem[Wu et~al.(2022)Wu, Chiu, and Yang]{wu2022jukedrummer}
Wu, Y.-K., Chiu, C.-Y., and Yang, Y.-H.
\newblock {JukeDrummer}: Conditional beat-aware audio-domain drum accompaniment
  generation via transformer {VQ-VAE}.
\newblock In \emph{ISMIR}, 2022.

\bibitem[Yeh et~al.(2020)Yeh, Hsiao, Fukayama, Kitahara, Genchel, Liu, Dong,
  Chen, Leong, and Yang]{Yeh2020}
Yeh, Y.-C., Hsiao, W.-Y., Fukayama, S., Kitahara, T., Genchel, B., Liu, H.-M.,
  Dong, H.-W., Chen, Y., Leong, T., and Yang, Y.-H.
\newblock Automatic melody harmonization with triad chords: A comparative
  study.
\newblock \emph{arXiv:2001.02360}, 2020.

\bibitem[Zeghidour et~al.(2021)Zeghidour, Luebs, Omran, Skoglund, and
  Tagliasacchi]{zeghidour2021soundstream}
Zeghidour, N., Luebs, A., Omran, A., Skoglund, J., and Tagliasacchi, M.
\newblock {SoundStream}: An end-to-end neural audio codec.
\newblock \emph{IEEE/ACM Transactions on Audio, Speech, and Language
  Processing}, 2021.

\end{thebibliography}
\bibliographystyle{icml2023}

\clearpage

\clearpage
\newpage
\appendix


\section{Additional experimental results}

\begin{table*}[t]
\vskip 0.15in
\centering
\raisebox{-1.0in}{\rotatebox{90}{\raisebox{.02in}{$\underset{\overbrace{\hspace{1.4in}}{}}{\texttt{Noisy}}$}\hspace{.05in}$\underset{\overbrace{\hspace{.35in}}{}}{\texttt{Clean}}$}}
\begin{tabular}{lcccccc}
\toprule
\tabapprow{Method}{\text{Steps}}{\fadi}{\fads}{$\Delta$}{\nlli}{\nlls}
\midrule
\tabapprow{\expname{SA-SA}}{$23$k}{$3.01$}{$1.61$}{$1.39$}{$3.84$}{$3.79$}
\tabapprow{\expname{S-SA}}{$148$k}{$2.31$}{$\mathbf{1.14}$}{$1.17$}{$3.72$}{$3.71$}
\midrule
\tabapprow{\expname{SA-SA}}{$18$k}{$2.01$}{$1.64$}{$0.37$}{$3.92$}{$3.89$}
\tabapprow{\expname{~~NoFilt}}{$20$k}{$3.33$}{$2.30$}{$1.03$}{$3.93$}{$3.88$}
\tabapprow{\expname{~~RelPos}}{$43$k}{$2.24$}{$1.07$}{$1.17$}{$3.80$}{$3.79$}
\tabapprow{\expname{A-A}}{$18$k}{$3.41$}{$3.30$}{$\mathbf{0.11}$}{$4.03$}{$3.99$}
\tabapprow{\expname{SA-A}}{$43$k}{$2.81$}{$1.87$}{$0.95$}{$3.94$}{$3.89$}
\tabapprow{\expname{A-SA}}{$48$k}{$2.01$}{$1.65$}{$0.36$}{$3.85$}{$3.79$}
\tabapprow{\expname{S-SA}}{$148$k}{$\mathbf{1.36}$}{$1.17$}{$0.19$}{$\mathbf{3.71}$}{$3.70$}
\midrule
\tabapprow{\expname{S-SA-XL}}{$150$k}{$\mathbf{1.28}$}{$\mathbf{0.96}$}{$0.32$}{$\mathbf{3.47}$}{$\mathbf{3.47}$}
\bottomrule
\end{tabular}
\caption{
Expanded quantitative evaluation for our experiments. \fad{} is the Fr\'echet Audio Distances on MUSDB18-test between ground trunth mixes and mixes of ground truth vocals and instrumentals generated from either isolated vocals (\fadi{}) or source-separated vocals (\fads{}) as model inputs. Since \fadi{} corresponds better to the anticipated usage, and \fads{} is closer to how the model is trained, a generalization gap ($\Delta$) exists between the two. \nll{} is the negative log likelihood over acoustic instrumental codes given source-separated vocals (\nlls{}) and isolated vocals (\nlli{}) as model inputs. Adding noise to the input vocals (\textit{Noisy}), removing the vocal acoustic tokens (\expname{S-SA}), increasing model size (\expname{XL}) all help to improve generation for isolated vocals (\fadi{}).
Full details of featurizations can be found in Section~\ref{sec:features}. For all experiments, \fad{} on MUSDB18-dev is used as an early stopping metric, and ``Steps'' is the number of training steps before early stopping.}
\label{tab:quant_verbose}
\vskip -0.1in
\end{table*}

\begin{figure}[t]
\begin{center}
\centerline{\includegraphics[width=0.98\columnwidth]{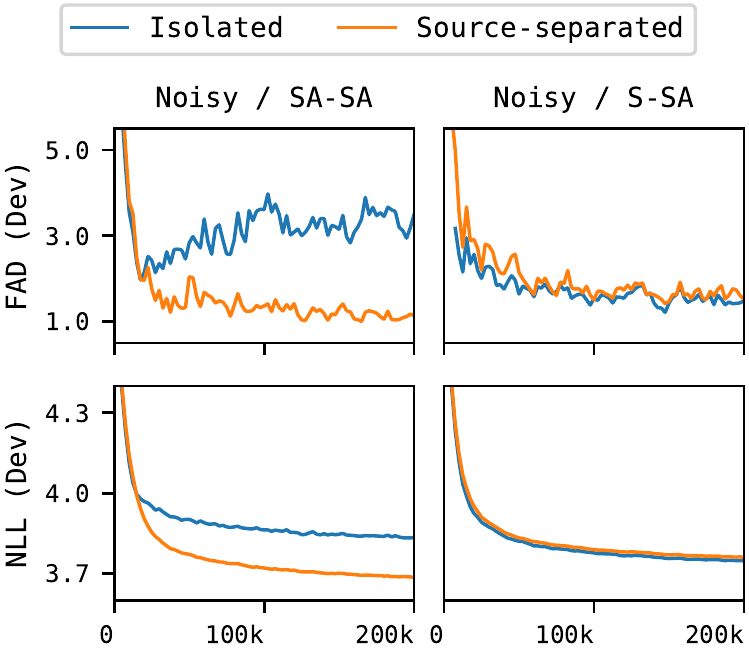}}
\caption{Compared to Frechet Audio Distance (FAD), we find negative log-likelihood (NLL) to be a poor indicator of subjective performance for this task.}
\label{fig:nllbad}
\end{center}
\end{figure}

In~\Cref{fig:nllbad}, we compare \fad{} and \nll{} for two of our experiments (\texttt{SA-SA} and \texttt{S-SA} under the \texttt{Noisy} condition) using both isolated and source-separated vocals as input.  
We note that, for both experiments, \nll{} on isolated vocals decreases monotonically, while in one experiment, \fad{} on isolated vocals diverges. 
Moreover, in listening to sound examples from different checkpoints, we found our subjective opinions to agree with \fad{} more often than with \nll{}. 
Hence, we center our quantitative evaluation and analysis around \fad{} in this work.
We report \nll{} on isolated (\nlli) and source-separated vocal inputs (\nlls) in~\Cref{tab:quant_verbose}). 

We also report results for two additional experiments, both under the \texttt{Noisy}~/~\texttt{SA-SA} condition:
(1)~\texttt{NoFilt} removes the dataset filtering described in~\Cref{sec:filter}, and 
(2)~\texttt{RelPos} uses relative positional embeddings---the default from T5~\citep{raffel2020exploring}---rather than the absolute positional encodings from vanilla Transformer~\citep{vaswani2017attention} that we use in our other experiments. 
We find that both removing filtering and using relative positional embeddings results in worse \fadi{} compared to using filtering and absolute positional encodings. 
Absolute positional encodings are also more computationally efficient especially for longer sequences.

\section{Additional listening study details}

\begin{figure}[ht]
\begin{center}
    \centerline{\includegraphics[clip, width=0.35\textwidth]{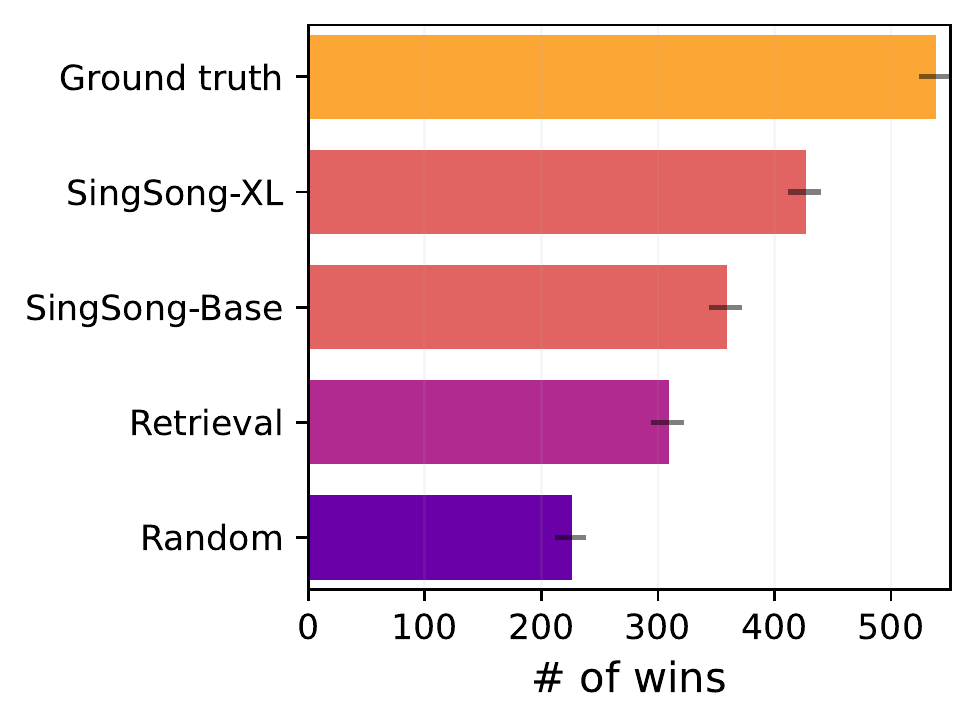}}
    \caption{Results of the human listener study. The number of wins is counting how often a model was preferred over another model in a side-by-side comparison, with error bars showing standard deviations in the mean.}
    \label{fig:listener-study}
\end{center}
\end{figure}

In~\Cref{fig:listener-study}, we show another view of the results of our listener study in terms of raw model ``wins'' on all pairwise comparisons against other systems ($N = 800$), with error bars showing standard deviation of estimated win proportion.

\begin{figure*}[ht]
\begin{center}
    \centerline{\includegraphics[clip, width=0.65\textwidth]{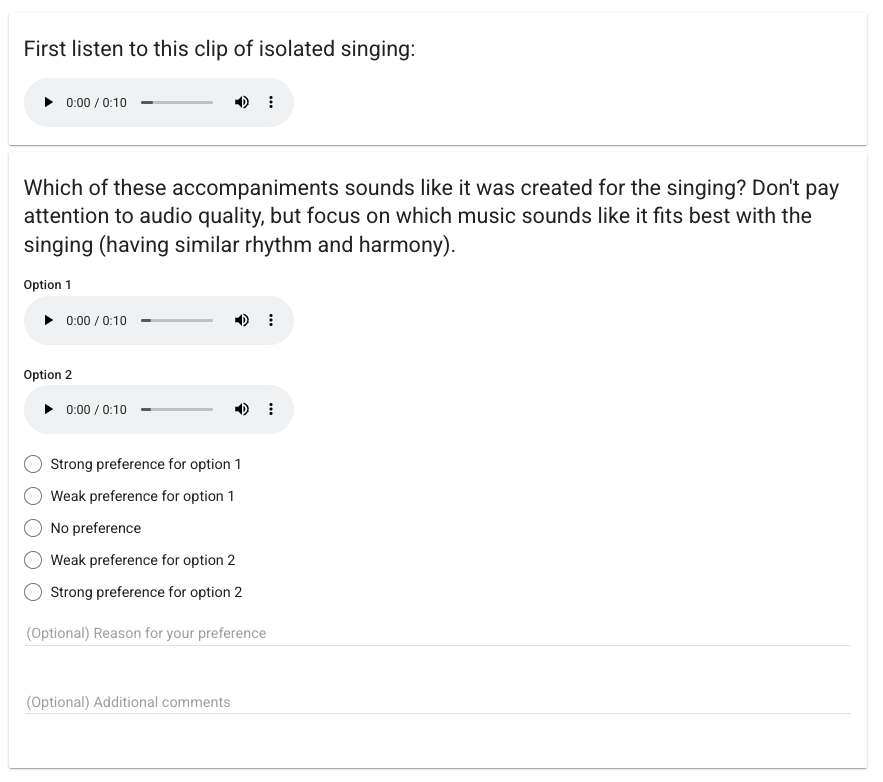}}
    \caption{A screenshot of the rating interface that participants of our listener study used to make their assessments. The audio clips labeled Option 1 \& Option 2 (below the prompt), are mixtures of the vocal input and accompaniment from one of the five methods we focus on: [\texttt{Ground Truth}, \texttt{SingSong-XL}, \texttt{SingSong-Base}, \texttt{Retrieval}, \texttt{Random}].}
    \label{fig:rater-ui}
\end{center}
\end{figure*}

~\cref{fig:rater-ui} shows the interface that participants in our listener study encountered when making their rating assessments. Raters were explicitly asked not to pay attention to audio quality, and rather focus on the best fit between the vocals and instrumental tracks. The audio clips that they rated were mixtures of the input vocals (top) and one of the five methods we study. Specifically, raters were prompted with an audio clip of isolated singing and presented with two mixtures of two of the accompaniment systems we test. The raters were then asked:

\begin{quote}
    Which of these accompaniments sounds like it was created for the singing? Don't pay attention to audio quality, but focus on which music sounds like it fits best with the singing (having similar rhythm and harmony).
\end{quote}

\subsection{Additional retrieval baseline details}

Here we report additional implementation details for the baseline \texttt{Retrieval} system. 
First, we sought to keep the ratio of estimated tempi of the vocals and instrumental in the interval $[0.5, 2.0]$ to minimize artifacts from time stretching. Thus, while the ratio was less than $0.5$, we multiplied by $2.0$ and while the ratio between was greater than $2.0$, we divided by $2.0$. Second, the vocal queries were always $10$ seconds exactly, therefore we removed 4 tracks from the MUSDB18-dev retrieval set that were shorter than $20$ seconds, the maximum possible time stretching ratio.

\end{document}